\documentstyle[aps,preprint,tighten]{revtex}

\newcommand{\sn}{{\rm sn}}
\newcommand{\cn}{{\rm cn}}
\newcommand{\dn}{{\rm dn}}

\newcommand{\nc}{{\rm nc}}

\newcommand{\uT}{u_{\Theta}}
\newcommand{\ut}{u_{\theta}}
\newcommand{\vT}{\varphi_{\Theta}}
\newcommand{\vt}{\varphi_{\theta}}
\newcommand{\bx}{{\bf x}}
\newcommand{\RR}{I\!\!R}

\newcommand{\bu}{{\bf u}}
\newcommand{\be}{{\bf e}}

\draft

\begin{document}

\title{\bf Semiclassical approximation to the partition function of
a particle in $D$ dimensions\footnote{UCLA/99/TEP/33}}
\author{C. A. A. de Carvalho$^{\star,\dag}$}
\address{Department of Physics and Astronomy, \\
University of California, Los Angeles, CA 90095-1547}
\author{R. M. Cavalcanti$^{\ddag}$}
\address{Instituto de F\'\i sica, Universidade de S\~ao Paulo, \\ 
C.P.~66318, S\~ao Paulo, SP 05315-970, Brasil}
\author{E. S. Fraga$^{\P,\S}$ and S. E. Jor\'as$^{\star\star,\parallel}$}
\address{Instituto de F\'\i sica, 
Universidade do Estado do Rio de Janeiro, \\ 
Rua S\~ao Francisco Xavier 524, Rio de Janeiro, RJ 20550-013, Brasil}
\date{\today}
\maketitle

\begin{abstract}

We use a path integral formalism to derive
the semiclassical series for the partition function of a 
particle in $D$ dimensions. We analyze in particular the case of attractive 
central potentials, obtaining explicit expressions for the 
fluctuation determinant and for the semiclassical 
two-point function in the 
special cases of the harmonic and singe-well quartic anharmonic 
oscillators. The specific heat of the latter is compared  
to precise WKB estimates. We conclude by discussing the possible 
extension of our results to field theories.

\end{abstract}



\section{Introduction}
\label{introduction}

As is well known \cite{feynman,schulman,kleinert,wiegel},
the partition function of a particle of mass $m$ 
interacting with a potential $V(\bx)$ and a thermal reservoir
at temperature $T$ can be written as a path integral 
($\beta=1/k_BT$):
\begin{mathletters}
\label{PI1}
\begin{equation}
\label{Z1}
Z(\beta)=\int_{\RR^D} d^D\bx_0\,\rho(\beta;\bx_0,\bx_0)\;,
\end{equation}
\begin{equation}
\rho(\beta;\bx_0,\bx_0)=\int_{\bx(0)=\bx_0}^{\bx(\beta\hbar)=\bx_0}
[{\cal D}\bx(\tau)]\, e^{-S[\bx]/\hbar}\;,
\end{equation}
\begin{equation}
S[\bx]=\int_0^{\beta\hbar} d\tau
\left[\frac{1}{2}\,m\left(\frac{d\bx}{d\tau}\right)^2+V(\bx)\right]\;.
\end{equation}
This path integral may be approximated in a number of 
ways: depending on the circumstances, one may resort to perturbation 
theory around the exactly soluble harmonic oscillator, variational 
estimates, or lattice Monte Carlo calculations (such techniques carry 
over to Quantum Statistical Field Theory, where free fields play the 
role of unperturbed uncoupled harmonic oscillators). Semiclassical 
techniques can also be used in approximating this integral. It is 
their virtues and shortcomings in applications to Statistical 
Mechanics that we intend to discuss.
\end{mathletters}

Semiclassical techniques have proven extremely important in the 
discussion of the transition from Quantum to Classical Mechanics 
\cite{gutzwiller,berry}. In the present context, however, we shall 
use them in the opposite sense: to systematically incorporate 
fluctuations (thermal and quantum) to a description that has one 
or more solutions of the ``Euclidean'' equations of motion 
as its starting point. (Heretofore, we call these solutions 
``trajectories'' or ``classical paths''.)
The Euclidean character is of crucial 
importance: first, it restricts the trajectories to be global minima 
of the Euclidean action \cite{footnote1} --- any others are 
exponentially suppressed; 
in addition, it leads to classical mechanics problems whose 
potencial is {\em minus} the physical one. Since we are 
interested in traces of operators, only closed trajectories  
will contribute. All this dramatically reduces the number of 
trajectories. In the specific examples of the harmonic and 
single-well quartic anharmonic oscillators, only one trajectory 
exists once the initial position and ``time-of-flight'' 
$\beta\hbar$ are fixed.

Thanks to the features described in the previous paragraph, 
in a recent paper \cite{semi} we were able to construct the full
semiclassical series for the partition function of a particle in 
one dimension from the mere knowledge of the trajectories. 
We obtained fluctuation determinants in a straightforward manner, 
by-passing the solution of the equivalent boundary-value problems, 
generated all the terms of the series in a systematic way, and 
could show that each term
has a non-perturbative character, corresponding to sums over
infinite subsets of perturbative graphs. Furthermore, we showed 
\cite{bsas} that the construction actually contains {\em all} 
the perturbative diagrams and many more. As an application of the 
method, we evaluated the ground-state 
energy and the specific heat of the single-well quartic anharmonic 
oscillator, 
achieving for the former good agreement with precise numerical
results \cite{vinette}, and for the latter a result which
has the correct high temperature limit, in contrast with the
one obtained via conventional perturbation theory around the minima
of the potential.

In this article, we present the $D$-dimensional generalization of the
method. Indeed, we are able to prove, as in the 
one-dimensional case, that it is 
possible to evaluate each term of the semiclassical series for the 
partition function using the classical path(s) as the only input. For 
the sake of simplicity, we concentrate on the case of attractive central
potentials. In such potentials, as will be shown below, the only
trajectories that contribute to the partition function
are the ones with zero angular momentum. The discussion 
for arbitrary potentials is left for an appendix. As 
examples, we consider the isotropic harmonic oscillator 
and the single-well quartic anharmonic oscillator; in particular,
we compute the specific heat of the latter in the lowest order
semiclassical approximation for a few values of the temperature and for 
$D=1$, 2 and 3.

The article is organized as follows: Section II presents the derivation 
of the semiclassical series for a generic potential 
in an arbitrary number of dimensions, and the explicit formulae
for the fluctuation determinant and the semiclassical two-point function in the 
particular case of 
attractive central potentials; Section III illustrates
these results in the cases of the harmonic and single-well
quartic anharmonic oscillators; 
Section IV presents our conclusions. 
In the Appendix, we show how to obtain the fluctuation determinant and 
the semiclassical two-point function in the case of an arbitrary potential in $D$ 
dimensions.


\section{The semiclassical expansion in statistical mechanics}    
\label{statistical}

\subsection{General formalism}

The procedure to generate a semiclassical series for $Z(\beta)$,
Eq.\ (\ref{PI1}),
was carried out in detail in Ref.\ \cite{semi} for the one-dimensional
case ($D=1$).
Here we shall only sketch its generalization for arbitrary $D$
(for a detailed discussion of the semiclassical expansion in
quantum mechanics using path integrals, see Refs.\ \cite{morette,mizrahi}).
The first step is to find the {\em minima} 
$\bx_c(\tau)$ of the Euclidean
action $S[\bx]$. They satisfy the Euler-Lagrange equation
\begin{equation}
\label{E-L}
m\ddot{\bx}_c-\nabla V(\bx_c)=0\;,
\end{equation} 
subject to the boundary conditions $\bx_c(0)=\bx_c(\beta\hbar)=\bx_0$;
for simplicity, we shall assume here that there is only one minimum.
The next step is to functionally expand the Euclidean action around it.
Writing $\bx(\tau)=\bx_c(\tau)+\bu(\tau)$, with
$\bu(0)=\bu(\beta\hbar)=0$, we have 
$S[\bx]=S[\bx_c]+S_2[\bu]+\delta S[\bu]$, where
\begin{mathletters}
\label{expS}
\begin{equation}
S_2[\bu]\equiv\frac{1}{2}\int_0^{\beta\hbar}d\tau\,u_i(\tau)
\left[-m\,\frac{d^2}{d\tau^2}\,\delta_{ij}+
\partial_i\partial_jV(\bx_c)\right]u_j(\tau)\;,
\end{equation}
\begin{equation}
\delta S[\bu]\equiv\int_0^{\beta\hbar}d\tau\,\delta V(\tau,\bu)
\equiv\int_0^{\beta\hbar}d\tau\sum_{n=3}^{\infty}\frac{1}{n!}\,
\partial_{i_1}\ldots\partial_{i_n}V(\bx_c)\,u_{i_1}(\tau)\ldots
u_{i_n}(\tau)\;;
\end{equation}
the indices $i,j,\ldots$ run from 1 to $D$, and repeated indices
are summed.
Inserting this decomposition of $S$ into (\ref{PI1}) and
expanding $e^{-\delta S/\hbar}$ in a power series 
yields the semiclassical expansion of $Z(\beta)$:
\end{mathletters}
\begin{equation}
\label{series}
Z(\beta)=\int_{\RR^D}d^D\bx_0\,e^{-S[\bx_c]/\hbar}
\int_{\bu(0)=0}^{\bu(\beta\hbar)=0}[{\cal D}\bu(\tau)]\,
e^{-S_2[\bu]/\hbar}\sum_{n=0}^{\infty}\frac{1}{n!}
\left(-\frac{\delta S[\bu]}{\hbar}\right)^n\;.
\end{equation}

The first term of the series corresponds to the quadratic
approximation to the partition function, which we denote
by $Z_2(\beta)$:
\begin{eqnarray}
Z_2(\beta)&\equiv&\int_{\RR^D}d^D\bx_0\,e^{-S[\bx_c]/\hbar}
\int_{\bu(0)=0}^{\bu(\beta\hbar)=0}[{\cal D}\bu(\tau)]\,
e^{-S_2[\bu]/\hbar}
\nonumber \\
&=&\int_{\RR^D}d^D\bx_0\,e^{-S[\bx_c]/\hbar}\,\Delta^{-1/2}\;,
\label{z2}
\end{eqnarray}
where $\Delta$ is the determinant of the fluctuation operator ${\cal F}$:
\begin{equation}
\Delta={\rm Det}\,{\cal F},\qquad{\cal F}_{ij}
=-m\,\frac{d^2}{d\tau^2}\,\delta_{ij}+
\partial_i\partial_jV(\bx_c)\;.
\end{equation}
The other terms of the series (\ref{series}) lead to integrals
of the type
\begin{equation}
\langle u_{i_1}(\tau_1)\ldots u_{i_k}(\tau_k)\rangle
\equiv\int_{\bu(0)=0}^{\bu(\beta\hbar)=0}[{\cal D}\bu(\tau)]\,
e^{-S_2[\bu]/\hbar}\,u_{i_1}(\tau_1)\ldots u_{i_k}(\tau_k)\;.
\end{equation}
Since the action $S_2[\bu]$ is quadratic, one can show that
\begin{equation}
\langle u_{i_1}(\tau_1)\ldots u_{i_k}(\tau_k)\rangle=
\hbar^{k/2}\,\Delta^{-1/2}\sum_P
G_{i_{j_1}i_{j_2}}(\tau_{j_1},\tau_{j_2})
\cdots G_{i_{j_{k-1}}i_{j_k}}(\tau_{j_{k-1}},\tau_{j_k})\;,
\end{equation}
if $k$ is even, and zero otherwise.
$\sum_P$ denotes the sum over all possible pairings of the $\tau_{j_k}$,
and $G_{ij}(\tau,\tau')$ is the solution of
\begin{equation}
\label{EqG}
\left[-m\,\frac{d^2}{d\tau^2}\,\delta_{ij}+
\partial_i\partial_jV(\bx_c)\right]G_{jk}(\tau,\tau')=
\delta_{ik}\,\delta(\tau-\tau')\;,
\end{equation}
satisfying the boundary conditions
\begin{equation}
\label{bc}
G_{jk}(0,\tau')=G_{jk}(\beta\hbar,\tau')=0.
\end{equation}
In the Appendix we present a recipe for obtaining 
$\Delta$ and $G_{ij}(\tau,\tau')$ using the general solution
of the equation of motion (\ref{E-L}) as the only input.


\subsection{Central potentials}

To illustrate the formalism of the previous section, let us
apply it to the case of central potentials, i.e., $V=V(r)$,
where $r\equiv|\bx|$. First of all, we note that,
because of the radial symmetry, 
$\rho(\beta;\bx_0,\bx_0)$ can only depend on $r_0=|\bx_0|$.
Thus, without loss of generality, we may take $\bx_0=r_0\,\be_1$,
where $\be_1$ is the unit vector pointing in the $x_1$-direction,
and perform the angular integration in (\ref{Z1}) to obtain
\begin{equation}
Z(\beta)=\frac{2\pi^{D/2}}{\Gamma(D/2)}\,\int_0^{\infty}dr_0\,r_0^{D-1}\,
\rho(\beta;r_0\,\be_1;r_0\,\be_1)\;.
\end{equation}
In general, there are many classical trajectories satisfying
the boundary conditions $\bx(0)=\bx(\beta\hbar)=r_0\,\be_1$. 
However, they are all radial if the potential
is purely attractive [i.e., $V'(r)>0$ for $r>0$]. Indeed,
in this case the Euclidean motion 
is equivalent to that of a particle in a repulsive 
central potential, so that a closed classical trajectory 
necessarily has zero angular momentum. Besides, this trajectory
is unique if the potential
is smooth at the origin, i.e., $V'(0)=0$.

For a trajectory lying in the $x_1$-axis,
$\bx_c(\tau)=r_c(\tau)\,\be_1$,
the fluctuation operator ${\cal F}$ is diagonal in the indices $i,j$.
Indeed, since $V=V(r)$, we have 
%
\begin{equation}
\partial_i\partial_jV(r)=\frac{V'(r)}{r}\,\delta_{ij}
+\left[V''(r)-\frac{V'(r)}{r}\right]\frac{x_ix_j}{r^2}\;,
\end{equation}
which, for $x_i=r_c\,\delta_{i1}$, gives
$\partial_1\partial_1V(r_c)=V''(r_c)$,
$\partial_i\partial_iV(r_c)=r_c^{-1}\,V'(r_c)$ for $i=2,\ldots,D$, and
$\partial_i\partial_jV(r_c)=0$ if $i\ne j$.
Thus, $\Delta=\Delta_{\ell}\,\Delta_t^{D-1}$, where 
\begin{equation}
\Delta_{\ell}={\rm Det}\,[-m\,\partial_{\tau}^2+V''(r_c)]\;,\qquad
\Delta_t={\rm Det}\,[-m\,\partial_{\tau}^2+r_c^{-1}\,V'(r_c)]
\end{equation}
($\ell$ and $t$ stand for {\em longitudinal} and {\em transverse},
respectively).

The Green's function $G_{ij}$ also becomes diagonal in this case:
$G_{11}=G_{\ell}$, $G_{ii}=G_t$ for $i=2,\ldots,D$, 
and $G_{ij}=0$ if $i\ne j$, where
\begin{mathletters}
\begin{equation}
[-m\,\partial_{\tau}^2+V''(r_c)]\,G_{\ell}(\tau,\tau')
=\delta(\tau-\tau')\;,
\end{equation}
\begin{equation}
[-m\,\partial_{\tau}^2+r_c^{-1}\,V'(r_c)]\,G_t(\tau,\tau')
=\delta(\tau-\tau')\;.
\end{equation}

$\Delta_{\ell}$ and $G_{\ell}(\tau,\tau')$ are the fluctuation
determinant and semiclassical Green's function that appear in the one-dimensional
version of the problem, which was studied in detail in 
Ref.\ \cite{semi}. There, the following results were 
derived:
%
\end{mathletters}
\begin{equation}
\label{DGl}
\Delta_{\ell}=\frac{2\pi\hbar}{m}\,\Omega_{\ell}(0,\beta\hbar)\;,
\qquad G_{\ell}(\tau,\tau')=\frac{\Omega_{\ell}(0,\tau_{<})\,
\Omega_{\ell}(\tau_{>},\beta\hbar)}
{m\,\Omega_{\ell}(0,\beta\hbar)}\;,
\end{equation}
where $\tau_{<}(\tau_{>})\equiv{\rm min}({\rm max})\{\tau,\tau'\}$
and
\begin{equation}
\label{Omegal}
\Omega_{\ell}(\tau,\tau')\equiv\frac{\eta_a(\tau)\,\eta_b(\tau')
-\eta_a(\tau')\,\eta_b(\tau)}{\eta_a(\tau')\,\dot{\eta}_b(\tau')
-\dot{\eta}_a(\tau')\,\eta_b(\tau')}\;,
\end{equation}
with $\eta_a(\tau)$ and $\eta_b(\tau)$ any two linearly independent
solutions of the homogeneus equation
\begin{equation}
\label{hl}
[-m\,\partial_{\tau}^2+V''(r_c)]\,\eta(\tau)=0.
\end{equation}
By differentiating the equation of motion $m\,\ddot{r}_c-V'(r_c)=0$
with respect to $\tau$, one can verify that 
$\eta_a(\tau)=\dot{r}_c(\tau)$ is one such solution. The other
can be taken as \cite{footnote2}
$\eta_b(\tau)=\dot{r}_c(\tau)\,\int_0^{\tau}d\tau'\,
[\dot{r}_c(\tau')]^{-2}$. For such a
choice, the denominator of $\Omega_{\ell}(\tau,\tau')$
is equal to 1 and, since $\eta_b(0)=0$, one has
$\Delta_{\ell}=(2\pi\hbar/m)\,\eta_a(0)\,\eta_b(\beta\hbar)$.
Because of these simplifying features, we shall refer to
those solutions as the ``canonical'' solutions of (\ref{hl}).

In order to obtain $\Delta_t$ and $G_t(\tau,\tau')$ one simply
replaces $\Omega_{\ell}(\tau,\tau')$ in (\ref{DGl}) by 
\begin{equation}
\label{Omegat}
\Omega_t(\tau,\tau')\equiv\frac{\varphi_a(\tau)\,\varphi_b(\tau')-
\varphi_a(\tau')\,\varphi_b(\tau)}{\varphi_a(\tau')\,\dot{\varphi}_b(\tau')
-\dot{\varphi}_a(\tau')\,\varphi_b(\tau')}\;,
\end{equation}
where $\varphi_a(\tau)$ 
and $\varphi_b(\tau)$ are two linearly independent solutions of
\begin{equation}
\label{ht}
[-m\,\partial_{\tau}^2+r_c^{-1}\,V'(r_c)]\,\varphi(\tau)=0\;.
\end{equation}
It immediately follows from the equation of motion that 
$\varphi_a(\tau)=r_c(\tau)$ is one such solution. Another one
is $\varphi_b(\tau)=r_c(\tau)\,\int_0^{\tau}d\tau'\,
[r_c(\tau')]^{-2}$. They form a pair of canonical solutions
of (\ref{ht}).


\section{Applications}

Using the results of the previous section we may write the
quadratic approximation to the partition function as
\begin{equation}
Z_2(\beta)=\frac{2\pi^{D/2}}{\Gamma(D/2)}\,\int_0^{\infty}dr_0\,
r_0^{D-1}\,e^{-S[\bx_c]/\hbar}\left(\Delta_{\ell}\,
\Delta_t^{D-1}\right)^{-1/2}\;.
\end{equation}
It can be readily calculated from the knowledge of 
$\bx_c(\tau)$ alone. This will be acomplished below
for both the harmonic and single-well quartic anharmonic oscillators.

\subsection{The harmonic oscillator}
\label{harmonic}

As a first example, we consider the $D$-dimensional (isotropic)
harmonic oscillator,
\begin{equation}
V(r)=\frac{1}{2}\,m\omega^2r^2\;.
\label{vhx}
\end{equation}
Since the potential is quadratic, $\delta V(\tau,\bu)=0$
and $Z(\beta)=Z_2(\beta)$. Besides, $r^{-1}\,V'(r)=V''(r)$,
so that $\Delta_t=\Delta_{\ell}$. Thus,
\begin{equation}
\label{ZHO}
Z(\beta)=\frac{2\pi^{D/2}}{\Gamma(D/2)}\,\int_0^{\infty}dr_0\,
r_0^{D-1}\,e^{-S[r_c]/\hbar}\,\Delta_{\ell}^{-D/2}\;.
\end{equation}
The solution of the equation of motion is straightforward, and yields
\begin{equation}
r_c(\tau)=\frac{r_0\,\cosh[\omega(\tau-\beta\hbar/2)]}
{\cosh(\beta\hbar\omega/2)}\;.
\end{equation}
The classical action can be readily computed, giving
\begin{equation}
\label{SHO}
S[r_c]=m\omega r_0^2\,\tanh(\beta\hbar\omega/2)\;.
\end{equation}
As solutions of (\ref{hl})
we may take $\eta_a(\tau)=\cosh(\omega\tau)$ and 
$\eta_b(\tau)=\sinh(\omega\tau)$. This gives $\Omega_{\ell}(\tau,\tau')
=\omega^{-1}\,\sinh[\omega(\tau'-\tau)]$, so that
\begin{equation}
\label{DHO}
\Delta_{\ell}=\frac{2\pi\hbar\,\sinh(\beta\hbar\omega)}{m\omega}\;.
\end{equation}
Inserting (\ref{SHO}) and (\ref{DHO}) into (\ref{ZHO})
and performing the integral, we obtain
\begin{equation}
Z(\beta)=\left[2\,\sinh(\beta\hbar\omega/2)\right]^{-D}\;,
\end{equation}
which is the well-known result for the partition function of the 
$D$-dimensional harmonic oscillator.


\subsection{The single-well quartic anharmonic oscillator}
\label{quartic}

Let us now consider the potential 
\begin{equation}
V(r)=\frac{1}{2}\,m\omega^2r^2+\frac{1}{4}\,\lambda r^4 \qquad(\lambda>0)\;.
\label{vx}
\end{equation}
In order to simplify notation, it is convenient to replace $r$ and $\tau$
by $q\equiv(\lambda/m\omega^2)^{1/2}\,r$ and $\theta\equiv\omega\tau$,
respectively. In the new variables, the equation of motion reads
\begin{equation}
\frac{d^2q}{d\theta^2}=q+q^3\;,
\label{ux}
\end{equation}
whose solution, taking into account the boundary conditions, is
\begin{equation}
q_c(\theta)=q_t\, \nc (u_{\theta},k)\;,
\label{classical}
\end{equation}
where $\nc (u,k)\equiv 1/\cn (u,k)$ is one of the Jacobian Elliptic 
functions \cite{grads,byrd,as}, and
\begin{equation}
u_{\theta}=\sqrt{1+q_t^2}\left(\theta-\frac{\Theta}{2}\right),
\qquad k=\sqrt{\frac{2+q_t^2}{2\,(1+q_t^2)}}\;,
\label{u,k}
\end{equation}
where $\Theta\equiv\beta\hbar\omega$.
The relation between $q_0$ and $q_t$ is obtained by taking
$\theta=\Theta$ in (\ref{classical}):
\begin{equation}
\label{q0qt}
q_0=q_c(\Theta)=q_t\,\nc\,\uT\;.
\label{q0-qt}
\end{equation}
(From now on we shall omit the $k$-dependence in the Jacobian Elliptic
functions.)

The classical action can be written as $S[r_c]=(m^2\omega^3/\lambda)\,
I[q_c]$, where
\begin{equation}
I[q]=\int_0^{\Theta}d\theta\left[\frac{1}{2}\,\dot{q}^2+U(q)\right],
\qquad U(q)=\frac{1}{2}\,q^2+\frac{1}{4}\,q^4\;.
\end{equation}
Using $\frac{1}{2}\,\dot{q}_c^2-U(q_c)=-U(q_t)$,
we may rewrite $I[q_c]$ as
\begin{equation}
I[q_c]=\Theta\,U(q_t)+2\int_{q_t}^{q_0}dq\,\sqrt{2\,[U(q)-U(q_t)]}\;.
\end{equation}
Performing the integration and using (\ref{q0qt}) yields
\begin{eqnarray}
I[q_c]&=&\Theta\left(\frac{1}{2}\,q_t^2+
\frac{1}{4}\,q_t^4\right)+\frac{4}{3}\left\{-\sqrt{1+q_t^2}
\left[{\rm E}(\vT,k)+\frac{1}{2}\,q_t^2\,\uT\right]\right.
\nonumber \\
& &+\left.\sn\,\uT\left(1+\frac{1}{2}\,q_t^2\,\nc^2\uT
\right)\sqrt{1+\frac{1}{2}\,q_t^2\,(1+\nc^2\uT)}\right\}\;,
\label{iqc}
\end{eqnarray}
where ${\rm E}(\varphi,k)$ denotes the Elliptic Integral of the 
Second Kind and
$\varphi_{\theta}\equiv\arccos[q_c(\theta)/q_0]=\arccos(\cn\,\ut)$.

The canonical solutions of (\ref{hl}) and (\ref{ht}) are given by
\begin{mathletters}
\begin{eqnarray}
\eta_a(\theta)&=&\omega q_t\,\sqrt{1+q_t^2}\,
\frac{\sn\,\ut\,\dn\,\ut}{\cn^2\ut}\;,
\\
\eta_b(\theta)&=&
\frac{1}{\omega^2q_t\,(1+q_t^2)}\,\frac{\sn\,\ut\,\dn\,\ut}{\cn^2\ut}
\left[\frac{k^2-1}{k^2}\,\ut+\frac{1-2k^2}{k^2}\,{\rm E}(\vt,k)\right.
\nonumber \\
& &\left.-\frac{\cn\,\ut\,\dn\,\ut}{\sn\,\ut}+(k^2-1)\,
\frac{\sn\,\ut\,\cn\,\ut}{\dn\,\ut}-(\theta\to 0)\right]\;,
\\
\varphi_a(\theta)&=&q_t\,\nc\,\ut\;,
\\
\varphi_b(\theta)&=&\frac{\nc\,\ut}{\omega k^2q_t\,\sqrt{1+q_t^2}}
\left[{\rm E}(\vt,k)+(k^2-1)\,\ut-(\theta\to 0)\right]\;.
\end{eqnarray}
Thus,
\end{mathletters}
\begin{mathletters}
\begin{eqnarray}
\Delta_{\ell}&=&\frac{4\pi\hbar}{m\omega}\,
\frac{\sn^2\uT\,\dn^2\uT}{\sqrt{1+q_t^2}\,\cn^4\uT}
\left[\frac{1-k^2}{k^2}\,\uT+\frac{2k^2-1}{k^2}\,{\rm E}(\vT,k)\right.
\nonumber \\
& &\left.+\frac{\cn\,\uT\,\dn\,\uT}{\sn\,\uT}+(1-k^2)\,
\frac{\sn\,\uT\,\cn\,\uT}{\dn\,\uT}\right]\;,
\\
\Delta_t&=&\frac{4\pi\hbar}{m\omega}\,
\frac{\nc^2\uT}{k^2\sqrt{1+q_t^2}}\left[{\rm E}(\vT,k)
+(k^2-1)\uT\right]\;.
\end{eqnarray}
\end{mathletters}

We now have all the necessary ingredients to compute the quadratic
approximation to $Z(\beta)$:
\begin{equation}
Z_2(\beta)=\frac{2\pi^{D/2}}{\Gamma(D/2)}\left(\frac{m\omega^2}
{\lambda}\right)^{D/2}\int_0^{\infty}dq_0\,q_0^{D-1}\,e^{-I[q_c]/g}
\left(\Delta_{\ell}\,\Delta_t^{D-1}\right)^{-1/2}\;,
\end{equation}
where $g\equiv\hbar\lambda/m^2\omega^3$. However, to perform the
integral over $q_0$ one must write $I[q_c]$,
$\Delta_{\ell}$ and $\Delta_t$ in terms of $q_0$. In view of 
Eq.\ (\ref{q0qt}), it is much simpler to change the variable
of integration from $q_0$ to $q_t$:
\begin{equation}
\label{z2.2}
Z_2(\beta)=\frac{2\pi^{D/2}}{\Gamma(D/2)}\left(\frac{m\omega^2}
{\lambda}\right)^{D/2}\int_0^{q_{\Theta}}dq_t\,
\left(\frac{\partial q_0}{\partial q_t}\right)_{\Theta}\,
(q_t\,\nc\,\uT)^{D-1}\,e^{-I[q_c]/g}
\left(\Delta_{\ell}\,\Delta_t^{D-1}\right)^{-1/2}\;,
\end{equation}
where $q_{\Theta}=\lim_{q_0\to\infty}\,q_t(q_0,\Theta)$.
The Jacobian $(\partial q_0/\partial q_t)_{\Theta}$ can
be obtained directly from (\ref{q0qt}) by differentiation
or, more simply, by using the identity \cite{semi}
\begin{equation}
\left(\frac{\partial q_0}{\partial q_t}\right)_{\Theta}=
\frac{m\omega}{4\pi\hbar}\,\frac{U'(q_t)\,\Delta_{\ell}}
{\sqrt{2\,[U(q_0)-U(q_t)]}}\;.
\end{equation}

As an application, we may use (\ref{z2.2}) to
calculate the specific heat of the $D$-dimensional single-well 
quartic anharmonic oscillator, given by
\begin{equation}
\label{C}
C=\beta^2\,\frac{\partial^2}{\partial\beta^2}\,\ln Z\;.
\end{equation}
Using the program MAPLE, we computed this 
expression for a few values of the temperature \cite{maple}. 
In Fig.\ \ref{fig1}, we present the results for $D=1$, 2 and 3. 
In Fig.\ \ref{fig2}, we compare the semiclassical approximation with: i)
the classical result, in which the partition
function is given by
\begin{equation}
Z_{\rm cl}(\beta)=\left(\frac{m}{2\pi\hbar^2\beta}\right)^{D/2}
\int d^Dx\,e^{-\beta V(\bx)};
\end{equation}
ii) with the lowest order WKB approximation, in which the energy levels
entering the expression $Z=\sum_n\,e^{-\beta E_n}$ are
given by the Bohr-Sommerfeld formula 
\begin{equation}
\oint\sqrt{2m\,[E_n-V(x)]}\,dx=\left(n+\frac{1}{2}\right)h
\qquad(n=0,1,2,\ldots);
\label{bohrsom}
\end{equation}
and iii) with the specific heat of the harmonic oscillator.


\section{Conclusions}
\label{conclusions}

The results of the previous sections confirm the findings of 
references \cite{semi,bsas}, and generalize them to arbitrary $D$. 
The semiclassical approach finds the minima of the Euclidean action 
and expands around them. As a result, it generates a series whose 
terms correspond to resummations of infinite numbers of perturbative 
graphs plus additional ones. Our calculations show that even the lowest order 
semiclassical estimates improve on perturbation theory at low
temperatures and, in contrast with it, correctly describe the
high temperature regime. 

The comparison with WKB estimates, done for the one-dimensional case, 
is particularly interesting. Such estimates approximate the values 
of the energy levels of the single-well anharmonic oscillator to a 
high precision if $g\equiv\hbar\lambda/m^2\omega^3$ is
small, even if we restrict 
ourselves to the lowest order WKB quantization condition, given by the
Bohr-Sommerfeld formula \cite{WKB}.
They were then used to compute the partition function  
by actually performing the sums over eigenstates numerically. 
Thus, the WKB results can be considered ``quasi-exact''. By
contrast, the
semiclassical approach directly approximates the {\em whole} sum. 
Its lowest order agrees well with the quasi-exact WKB result at both 
high {\em and} low temperatures. At high $T$, this agreement just 
reflects the convergence of both results to the classical limit, 
something which is completely missed by perturbation theory. Only 
in the intermediate region does our result differ from WKB, although 
we expect this to be modified with the inclusion of next-to-leading 
orders. It is less accurate, as it approximates the whole sum, 
whereas WKB approximates each term in the sum; but it does 
incorporate and improve upon the virtues of perturbation theory 
at low temperatures, and of the classical limit at high ones. 
Results for $D=2$ and $D=3$ do follow the same pattern, although 
we have not compared them to WKB estimates.

The advantage of this method is that it reduces the whole quantum 
problem to the computation of (few) classical paths. From then on, 
a systematic procedure takes care of generating each term in the 
series. Paradoxically, this may also be its weakness: there
are systems for which the action does not have a global minimum,
but which are perfectly well-defined quantum-mechanically.
The Coulomb potential is a good example; there, depending on the 
values of $\beta$ and $r_0$, the number of classical paths may be two, 
one or zero. Besides, only in the two-solution regime do we have 
minima. Even then, they are local, not global ones. Therefore, our 
starting point seems ill-defined. This should not come as a surprise, 
however, since here the classical limit itself is ill-defined, as 
the potential is unbounded below. As a matter of fact, even the
usual time-slicing prescription to calculate the path integral
must be modified in the case of the Coulomb potential \cite{kleinert}.
Cases like this will require 
special consideration, although there exist suggestions in the 
literature as to how to treat similar situations of absence of 
classical paths in Quantum Mechanics \cite{mevMev}.
Nevertheless, we expect the techniques presented here to be useful in any 
problem which can be reduced to the calculation of partition 
or correlation functions in equilibrium Statistical Mechanics, 
as long as it allows for a simple 
analysis of the minima of the Euclidean action. 

Our next step is to investigate how the semiclassical treatment 
affects field-theoretic problems at finite temperature, 
where standard methods 
of computation of effective potentials rely on expansions around 
constant backgrounds. At finite temperature, these are not in 
general minima of the Euclidean action. This might lead to problems 
with the expansions around such backgrounds at high temperatures, 
of the same nature of those encountered by perturbation theory in 
Quantum Statistical Mechanics. Even if we neglect any coordinate 
dependence of the fields, their dependence on Euclidean 
time is essential to satisfy equations of motion and boundary 
conditions that caracterize classical paths. We expect this to 
have an effect on a variety of calculations.

Another problem of interest is to generalize our results to field 
theories with spherically symmetric classical solutions.  
An extension of the approach presented in this work to treat 
models containing non-trivial backgrounds (like instantons, 
monopoles, vortices, etc.) as classical solutions might lead 
to some new insights. Unfortunately, the extension of our results 
to field theories is not a straightforward process. In fact, we 
do not know how to construct a semiclassical propagator in general. 
The success of our program will depend on how well can we 
circumvent this difficulty.
       

\acknowledgements

The authors acknowledge support from CNPq, FAPERJ, FAPESP and
FUJB/UFRJ. CAAC would like to thank UCLA for its hospitality. 
ESF would like to thank Rob Pisarski 
and Larry McLerran for their kind hospitality at BNL, where part of 
this work has been done. ESF is partially supported by the U. S. 
Department of Energy under Contract No. DE-AC02-98CH10886 and by 
CNPq through a post-doctoral fellowship.


\appendix

\section{}
\label{secjacobi}

Let $J(\tau,\tau')$ be the solution of the homogeneous
differential equation
\begin{equation}
\left[-m\,\frac{d^2}{d\tau^2}\,\delta_{ij}+
\partial_i\partial_jV(\bx_c)\right]J_{jk}(\tau,\tau')=0
\label{flucthom}
\end{equation}
satisfying the initial conditions
\begin{equation}
J(\tau',\tau')=0\;,\qquad 
\frac{\partial}{\partial\tau}\,J(\tau=\tau',\tau')
=-\frac{1}{m}\,\openone\;.
\label{ic}
\end{equation} 
This function is known as the Jacobi 
commutator \cite{morette,mizrahi}.
It can be explicitly constructed as follows.
Let $\bx(\tau;{\bf a},{\bf b})$ be the solution of
the equation of motion (\ref{E-L}) satisfying the 
initial conditions $\bx(0)={\bf a}$, $\dot{\bx}(0)={\bf b}$.
Let $A$ and $B$ be the $D\times D$ matrices
defined as
\begin{equation}
A_{jk}(\tau)=\frac{\partial}{\partial a_k}\,
x_j(\tau;{\bf a}=\bx_0,{\bf b}={\bf v}_0), 
\end{equation}
\begin{equation}
B_{jk}(\tau)=\frac{\partial}{\partial b_k}\,
x_j(\tau;{\bf a}=\bx_0,{\bf b}={\bf v}_0),
\end{equation}
where ${\bf v}_0=\dot{\bx}_c(0)$. By differentiating 
Eq.\ (\ref{E-L}) with respect to $a_k$ and $b_k$
(and taking ${\bf a}=\bx_0$, ${\bf b}={\bf v}_0$),
one can show that they are solutions of (\ref{flucthom}).
They are also invertible for $\tau$ small 
enough \cite{footnote4} (but not zero). Indeed,
$\bx(\tau)={\bf a}+{\bf b}\tau+O(\tau^2)$ when $\tau\to 0$,
hence $A(\tau)=\openone+O(\tau^2)$ and 
$B(\tau)=\tau\,\openone+O(\tau^2)$.
Therefore, the expression
\begin{equation}
\label{Jtt'}
J(\tau,\tau')=-\frac{1}{m}\left[A(\tau)\,A^{-1}(\tau')
-B(\tau)\,B^{-1}(\tau')\right]
\left[\dot{A}(\tau')\,A^{-1}(\tau')
-\dot{B}(\tau')\,B^{-1}(\tau')\right]^{-1}
\end{equation}
makes sense, and one can easily verify that it
satisfies (\ref{flucthom}) and (\ref{ic}).

The Green's function $G(\tau,\tau')$ 
can be written in terms of the Jacobi commutator as 
\begin{equation}
\label{GJ}
G(\tau,\tau')=J(\tau,0)\,M(0,\beta\hbar)\,J(\beta\hbar,\tau')\,
\theta(\tau'-\tau) 
-J(\tau,\beta\hbar)\,M(\beta\hbar,0)\,
J(0,\tau')\,\theta(\tau-\tau')\;,
\end{equation}
where $M(\tau,\tau')=-J(\tau',\tau)^{-1}$ and $\theta(\tau)$ is the
Heaviside step function. 
To prove (\ref{GJ}) we need the following identities: 
\begin{equation}
J(\tau,0)\,M(0,\beta\hbar)\,J(\beta\hbar,\tau')
+J(\tau,\beta\hbar)\,M(\beta\hbar,0)\,J(0,\tau')=-J(\tau,\tau')\;,
\label{id1}
\end{equation}
\begin{equation}
\partial_{\tau}J(\tau,0)\,M(0,\beta\hbar)\,J(\beta\hbar,\tau)
+\partial_{\tau}J(\tau,\beta\hbar)\,M(\beta\hbar,0)\,J(0,\tau)
=\frac{1}{m}\,\openone\;.
\label{id2}
\end{equation}
The first identity follows from the fact that both functions
are solutions of the same second order differential equation 
[Eq.\ (\ref{flucthom})] and are equal at $\tau=0$
and $\tau=\beta\hbar$; the second follows from
(\ref{ic}) and (\ref{id1}).

Now, the proof of (\ref{GJ}): (i) it is a solution of (\ref{EqG})
when $\tau<\tau'$ or $\tau>\tau'$;
(ii) it satisfies the boundary conditions (\ref{bc}); (iii) it is
continuous at $\tau=\tau'$,
\begin{equation}
G(\tau'+0,\tau')=G(\tau'-0,\tau')
\end{equation}
[use (\ref{id1}) with $\tau=\tau'$],
and (iv) its derivative with respect to $\tau$ has the discontinuity
implied by (\ref{EqG}),
\begin{equation}
\frac{\partial}{\partial\tau}\,G(\tau=\tau'+0,\tau')
-\frac{\partial}{\partial\tau}\,G(\tau=\tau'-0,\tau')
=-\frac{1}{m}\,\openone
\end{equation}
[use (\ref{id2})].

Finally, the determinant $\Delta$ of the fluctuation operator 
${\cal F}$ is given by 
\begin{equation}
\Delta = (2\pi \hbar)^{D}\,{\rm det}\,[-J(\beta\hbar,0)]\;.
\end{equation}
This formula can be proven along the lines of
Appendix 1 of Ref.\ \cite{coleman}.


\begin{figure}
\caption{Specific heat (in units of $k_B$) 
vs.\ temperature ($T\equiv 1/\beta\hbar\omega$)
for the one- (diamonds), two- (circles), and three-dimensional (crosses)
single-well quartic anharmonic oscillator in the semiclassical approximation. 
$g\equiv\hbar\lambda/m^2\omega^3=0.5$.}
\label{fig1}
\end{figure}

\begin{figure}
\caption{Specific heat (in units of $k_B$) vs.\ temperature 
($T\equiv 1/\beta\hbar\omega$) 
for the one-dimensional harmonic oscillator (long-dashed line)
and for the single-well quartic anharmonic oscillator: classical result
(short-dashed line), semiclassical approximation (circles), and
WKB approximation (solid line). 
$g\equiv\hbar\lambda/m^2\omega^3=0.2$.}
\label{fig2}
\end{figure}

\begin{references}

\bibitem[\star]{CAAC1} On leave from {\sl Instituto de F\'\i sica,
Universidade Federal do Rio de Janeiro, C. P. 68528, Rio de Janeiro, 
RJ 21945-970, Brasil}

\bibitem[\dag]{CAAC2} E-mail: aragao@physics.ucla.edu

\bibitem[\ddag]{RMC} E-mail: rmoritz@fma.if.usp.br

\bibitem[\P]{ESF1} Present address: {\sl Nuclear Theory Group, 
Physics Department, Brookhaven National Laboratory, Upton, NY 11973-5000}

\bibitem[\S]{ESF2} E-mail: fraga@bnl.gov

\bibitem[\star\star]{SEJ1} Present address: {\sl High Energy Theory Group, 
Physics Department, Brown University, Providence, RI 02912}

\bibitem[\parallel]{SEJ2} E-mail: joras@het.brown.edu

\bibitem{feynman} R. P. Feynman and A. R. Hibbs, {\it Quantum Mechanics
and Path Integrals\/} (McGraw-Hill, New York, 1965); 
R. P. Feynman, {\it Statistical Mechanics\/} (Addison-Wesley, New York, 
1972).

\bibitem{schulman} L. S. Schulman, {\it Techniques and Applications of 
Path Integration} (John Wiley, New York, 1981).

\bibitem{kleinert} H. Kleinert, {\it Path Integrals in Quantum 
Mechanics, Statistics and Polymer Physics} (World Scientific, 
Singapore, 1995).

\bibitem{wiegel} F. W. Wiegel, {\it Introduction to Path-Integral Methods
in Physics and Polymer Science} (World Scientific, Singapore, 1986).

\bibitem{gutzwiller} M. C. Gutzwiller, {\it Chaos in Classical and 
Quantum Mechanics} (Springer-Verlag, New York, 1990).

\bibitem{berry} M. Berry,
{\it Some quantum-to-classical asymptotics},
in {\it Chaos and Quantum Physics}, edited by 
M.-J. Giannoni, A. Voros, and J. Zinn-Justin 
(North-Holland, Amsterdam, 1991).

\bibitem{footnote1} This is to be
contrasted with the semiclassical approximation in Quantum Mechanics,
where one has to find all the stationary points of the action, 
irrespective of being minima, maxima or points of inflection.

\bibitem{semi} C. A. A. de Carvalho, R. M. Cavalcanti, E. S. Fraga, and 
S. E. Jor\'as, Ann. Phys. (N.Y.) {\bf 273}, 146 (1999).

\bibitem{bsas} C. A. A. de Carvalho and R. M. Cavalcanti, in 
{\it Trends in Theoretical Physics II} (AIP Conference Proceedings
484), edited by H. Falomir, 
R. E. Gamboa Sarav\'{\i}, and F. A. Schaposnik (AIP, Woodbury, 1999); 
quant-ph/9903028.

\bibitem{vinette} F. Vinette and J. \v{C}\'{\i}\v{z}ek, J. Math. Phys.
{\bf 32}, 3392 (1991); W. Janke and H. Kleinert, Phys. Rev. Lett. {\bf 75},
2787 (1995).

\bibitem{morette} C. DeWitt-Morette, Comm. Math. Phys. {\bf 28}, 47 
(1972); {\bf 37}, 63 (1974); Ann. Phys. (N.Y.) {\bf 97}, 367 (1976).

\bibitem{mizrahi} Maurice M. Mizrahi, J. Math. Phys. {\bf 17}, 566 
(1976); {\bf 19}, 298 (1978); {\bf 20}, 844 (1979).

\bibitem{grads} I. S. Gradshteyn and I. M. Ryzhik, {\it Table of  
Integrals, Series, and Products\/} (Academic Press, New York, 1965).

\bibitem{byrd} P. F. Byrd and M. D. Friedman, {\it Handbook of Elliptic 
Integrals for Engineers and Physicists} (Springer-Verlag, Berlin, 1954).

\bibitem{as} M. Abramowitz and I. A. Stegun (eds.), {\it Handbook of
Mathematical Functions\/} (Dover, New York, 1965).

\bibitem{footnote2} This definition only makes sense for
$\tau<\beta\hbar/2$. For $\tau>\beta\hbar/2$ one must take
$$
\eta_b(\tau)=\dot{r}_c(\tau)\left(C+\int_{\beta\hbar}^{\tau}
\frac{d\tau'}{\dot{r}_c^2(\tau')}\right),
$$ 
with $C$ chosen in such a way
that $\eta_b(\tau)$ and $\dot{\eta}_b(\tau)$ are continuous
at $\tau=\beta\hbar/2$. See \cite{semi} for details.

\bibitem{maple} In order to compute the derivative in (\ref{C}),
we have approximated $\ln Z$ by a rational function using
Thiele's interpolation formula. See \cite{as}, formula 25.2.50.

\bibitem{WKB} For instance, when $g=0.2$ Eq.\ (\ref{bohrsom})
gives values of $E_0$, $E_1$ and $E_9$ that differ from the
exact ones by less than $3\%$, $1\%$ and $0.1\%$, respectively.

\bibitem{mevMev} L. S. Schulman and R. W. Ziolkowski, {\it Path integral asymptotics in the absence of classical paths}, in {\it Path Integrals 
from mev to Mev}, edited by V. Sa-yakanit et al (World Scientific, Singapore, 1989).

\bibitem{footnote4} We conjecture that they are in fact invertible
for any $\tau\in(0,\beta\hbar)$, except maybe for a finite 
number of points.

\bibitem{coleman} S. Coleman, {\it The uses of instantons}, in
{\it The Whys of Subnuclear Physics}, edited by A. Zichichi 
(Plenum, New York, 1979);
reproduced in S. Coleman, {\it Aspects of Symmetry} (Cambridge
University Press, Cambridge, 1985).

\end{references}
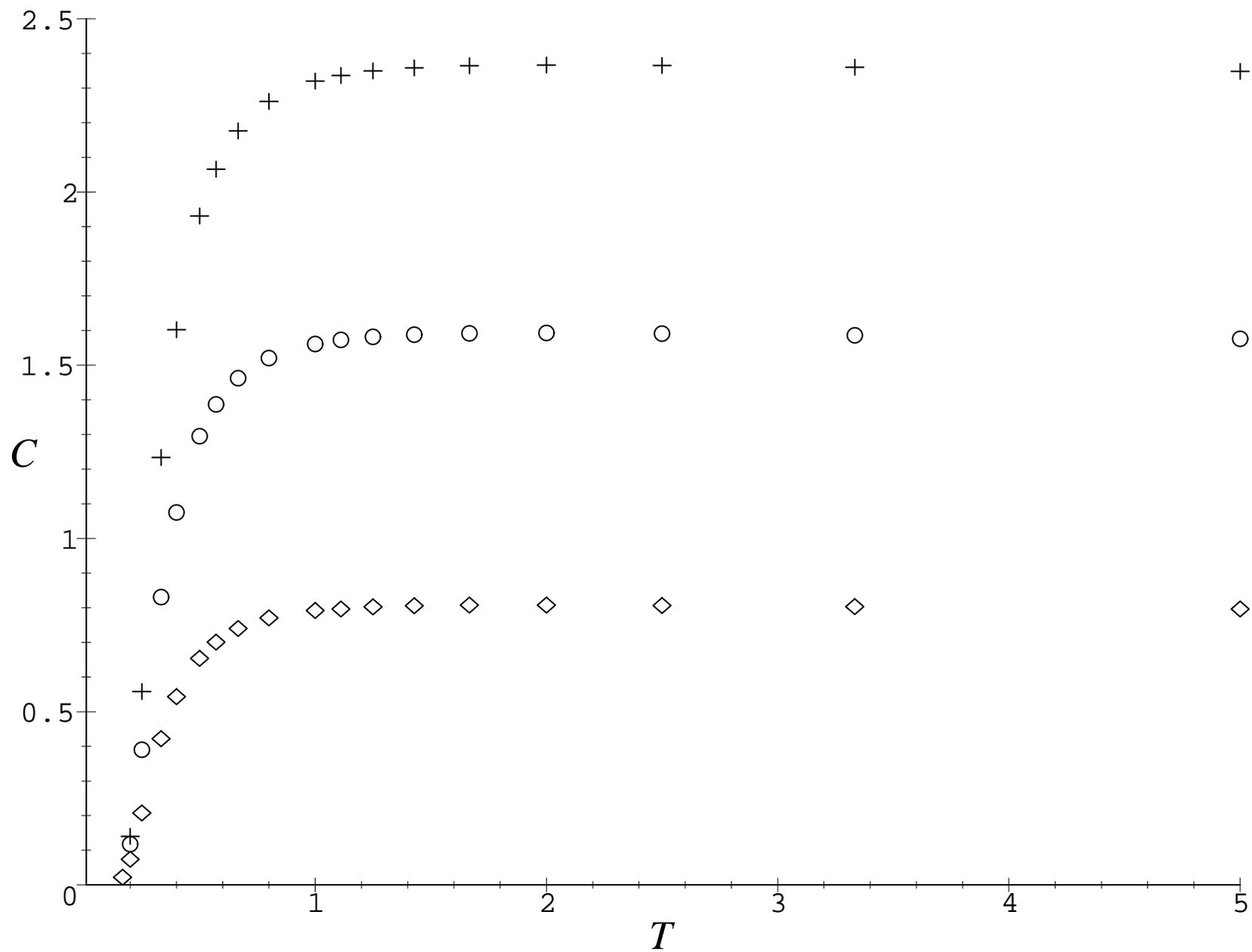
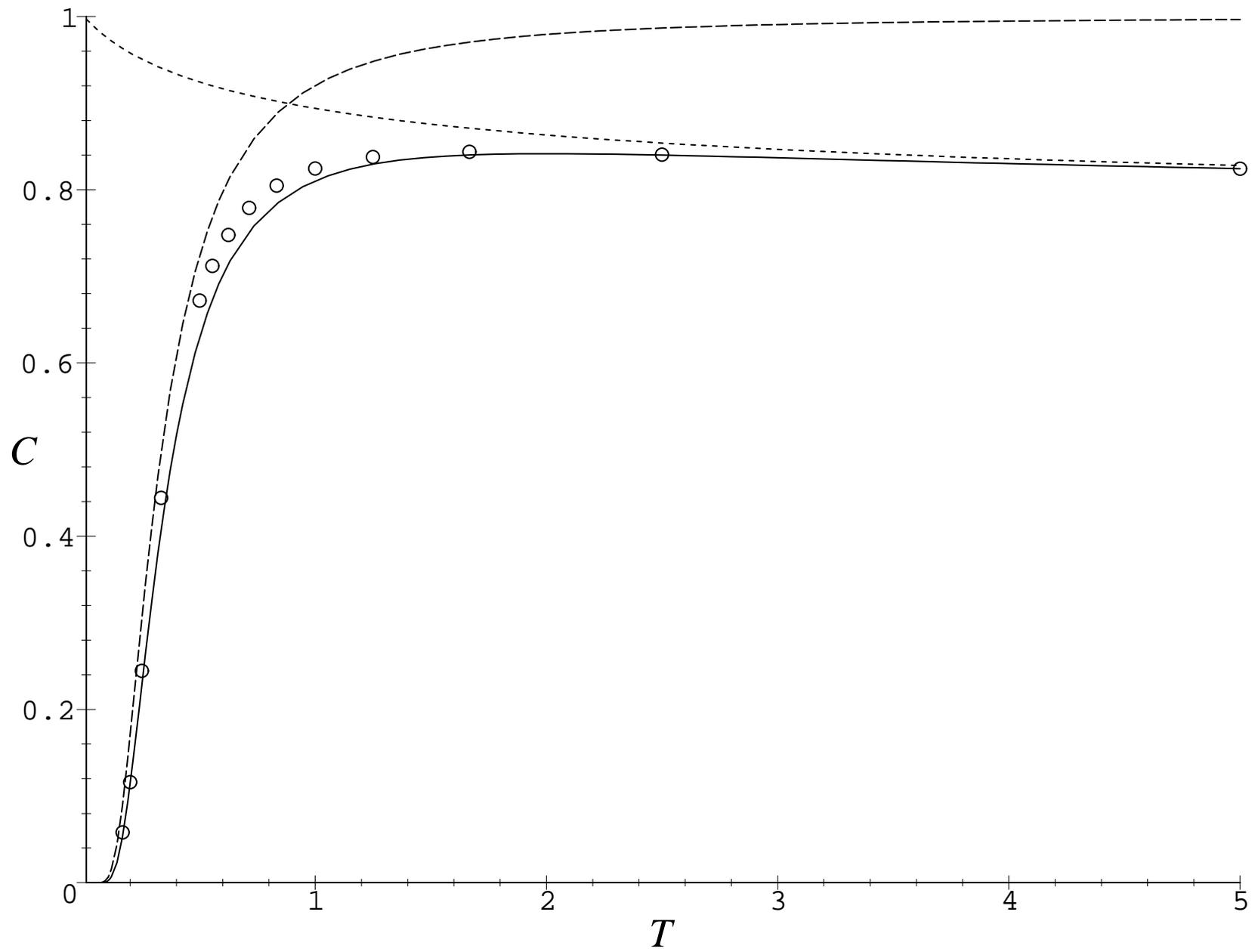
\end{document}